\documentclass{iau} 
\usepackage[pdftex]{graphicx}
\usepackage{natbib}
\usepackage{sidecap,caption}
\title[Multiple AGN activity in the XDCP0044 cluster core] 
{The role of AGN activity in the building up of the BCG at $z\sim1.6$}

\author[Angela Bongiorno \& Andrea Travascio]   
{Angela Bongiorno$^1$
 \and Andrea Travascio$^1$}

\affiliation{$^1$ INAF-Observatory of Rome, via di Frascati 33, \\ 00074, Monteporzio Catone, Rome, Italy\\ email: {\tt angela.bongiorno@inaf.it, andrea.travascio@inaf.it}}

\pubyear{2019}
\volume{356}  
\jname{IAUS 356: Nuclear Activity in Galaxies Across Cosmic Time}
\editors{M. Pov\'ic, P. Marziani, J. Masegosa, H. Netzer, \\
S. H. Negu \& S. B. Tessema, eds.}
\begin{document}

\maketitle

\begin{abstract}
XDCPJ0044.0-2033 is one of the most massive galaxy cluster at $z\sim1.6$, for which a wealth of multi-wavelength photometric and spectroscopic data have been collected during the last years. I have reported on the properties of the galaxy members in the very central region ($\sim70kpc\times70kpc$) of the cluster, derived through deep HST photometry, SINFONI and KMOS IFU spectroscopy, together with Chandra X-ray, ALMA and JVLA radio data. \\
In the core of the cluster, we have identified two groups of galaxies (Complex A and Complex B), seven of them confirmed to be cluster members, with signatures of ongoing merging. These galaxies show perturbed morphologies and, three of them show signs of AGN activity. In particular, two of them, located at the center of each complex, have been found to host luminous, obscured and highly accreting AGN ($\lambda=0.4-0.6$) exhibiting broad H$\alpha$ line. Moreover, a third optically obscured type-2 AGN, has been discovered through BPT diagram in Complex A. The AGN at the center of Complex B is detected in X-ray while the other two, and their companions, are spatially related to radio emission. The three AGN provide one of the closest AGN triple at z$>$1 revealed so far with a minimum (maximum) projected distance of 10 kpc (40 kpc).
The discovery of multiple AGN activity in a highly star-forming region associated to the crowded core of a galaxy cluster at $z\sim1.6$, suggests that these processes have a key role in shaping the nascent Brightest Cluster Galaxy, observed at the center of local clusters. According to our data, all galaxies in the core of XDCPJ0044.0-2033 could form a BCG of $M_{\star}\sim10^{12}M_\odot$ hosting a BH of $2\times10^{8}-10^{9}M_\odot$, in a time scale of the order of 2.5 Gyrs.

\end{abstract}

\firstsection 
\section{Introduction}

Relaxed, virialized and undisturbed galaxy cluster in the local Universe are characterised by a bright, massive and large elliptical galaxy at their center, the so called Brightest Cluster Galaxy (BCG). The BCG is usually located at the center of the cluster potential well, close to the peak of X-ray emission. How these galaxies form is still a matter of study.
According to most models, the epoch of their assembly is mostly between z=1 and z=2, where the mass of the BCG goes from 10\% to 50\% of the final mass. For this reason, this redshift range is crucial to observe the BCG progenitors and to witness its assembly.
Indeed, the cores of galaxy clusters at $z=1-2$  show a different picture compared to the local Universe, i.e. in most cases there is no single BCG, and the core is characterised by several galaxies which are typically blue, star forming and with disturbed morphology. This implies the existence of a mechanism able to drive such transformation. Interestingly, also looking at the properties of the whole galaxy cluster population, there are evidences that z=1.5 is a crucial epoch. Indeed, while at $z<1.4$ the number of star forming galaxies (SFGs) increases towards the cluster outskirt, at $z>1.4$ the SF activity is higher in the core of the cluster \citep{Brodwin13}. 
For all these reasons, we studied the core of the X-ray detected galaxy cluster XDCP J0044.0-2033 \citep[hereafter XDCP0044;][]{Santos11,Fassbender11} at $z\sim1.6$. 
XDCP0044 is the most massive galaxy clusters ($M_{200}=4\times10^{14}M_{\odot}$) discovered in the XDCP project  through XMM archival data \citep{Fassbender11} and one of the most massive at $z>1.5$. It is in a quite advanced state of dynamical relaxation, and shows a reversal SF-density relation \citep{Santos15}. XDCP0044 is a unique laboratory to study the building-up of the BCG and the interplay between galaxies, nuclear activity, and the inter-galactic gas in the core of massive high redshift galaxy clusters. 

\begin{figure}
\centering
\includegraphics[width=13 cm, angle=0]{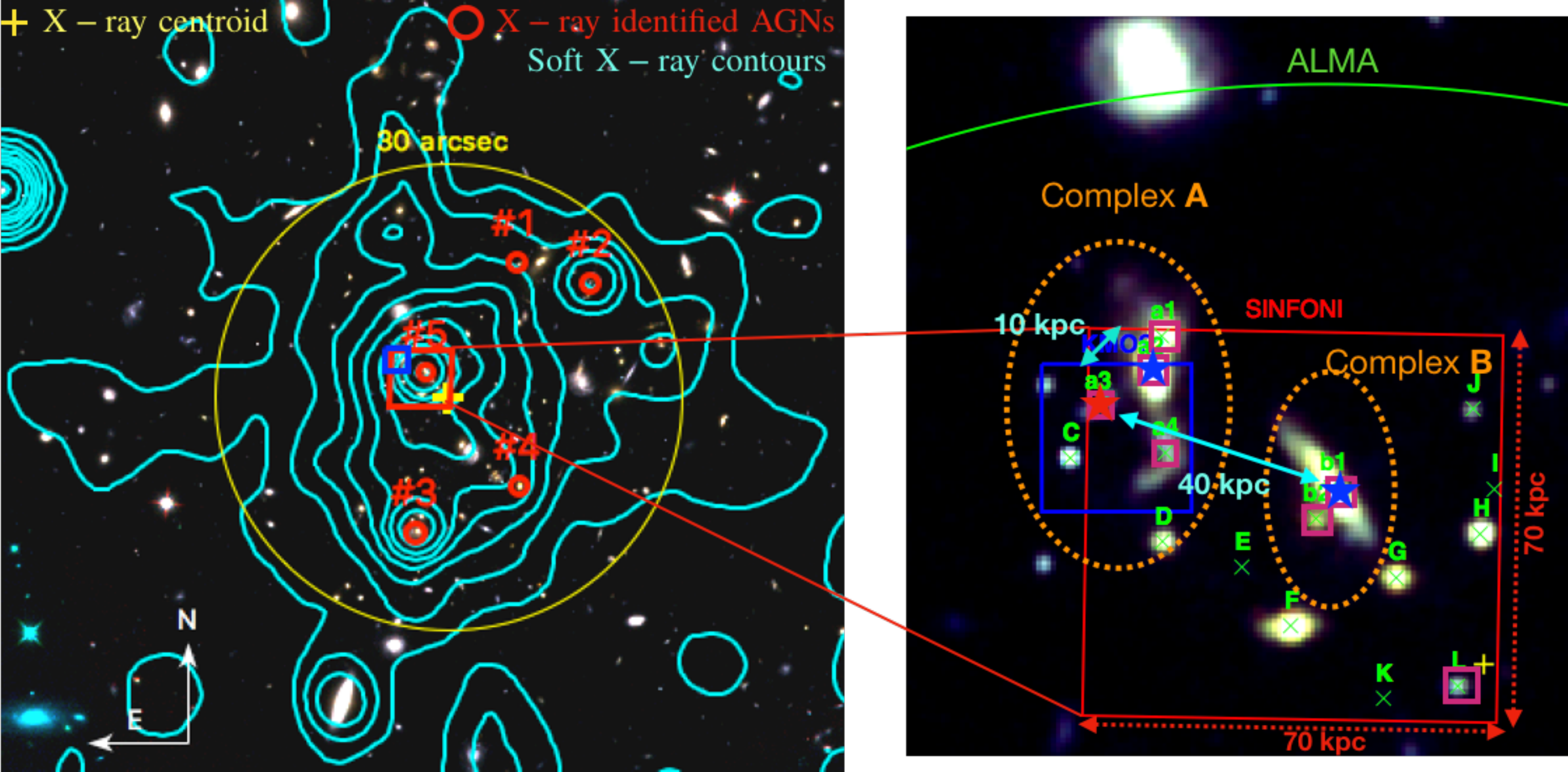}
\caption{[Adapted from \citealt{Travascio20}]\textit{Left Panel:}: HST RGB (F105W + F140W + F160W) image of the galaxy cluster \texttt{XDCP0044}. The yellow cross indicates the centroid of the extended X-ray emission while the yellow circle is $R_{2500}=250 kpc$. The cyan contours are the Chandra soft $ [0.5-2] keV$ X-ray emissions while the red circles mark the unresolved X-ray sources as identified by \citet{Tozzi15}. Finally, the red and blue squares delimit the region analyzed in this paper, corresponding to the SINFONI and the KMOS FOVs. \textit{Right Panel}: zoom-in of the analyzed central region, where two complexes (A and B) are highlighted with the orange dashed line. Green crosses mark the 16 HST 
identified photometric sources while magenta squares mark the sources for which a spectroscopic redshift has been determined. AGN are indicated with blue (type-1) and red (type-2) stars.}
\centering
\label{fig:HSTzoom}
\end{figure}

Fig. \ref{fig:HSTzoom} (left panel) shows the HST RGB (F105W+F140W+F160W) image of XDCP0044 with overlaid  the soft ($[0.5-2] keV$, cyan) 
 band X-ray Chandra contours. Red circles indicate the 5 point-like sources (AGN) identified by \citet{Tozzi15} within 30'' ($250 kpc$) from the cluster center.

In this work, we focused on a very small region ($70kpc\times70kpc$) of the cluster core (right panel of Fig. \ref{fig:HSTzoom}) for which a detailed multiwavelength study has been conducted  by combining the information derived from X-ray to optical, near-infrared (NIR) and radio bands, both photometrically and spectroscopilly \citep[see][]{Travascio20}.



\section{Results}
As visible in the right panel of Fig. \ref{fig:HSTzoom}, the core of XDCP0044, a very small region of 70kpc $\times$70kpc (slightly more than twice the milky way in size), appears very crowded. Through a SExtractor analysis  \citep{Bertin96},  we indeed distinguished 16 photometric sources (green crosses in Fig.\ref{fig:HSTzoom}, right panel) and thanks to the spectroscopic SINFONI and KMOS data, we confirmed that at least seven of them are cluster members (magenta squares in Fig.\ref{fig:HSTzoom}, right panel), with redshifts ranging from $z=1.5567$ to $z=1.5904$ ($\Delta z\simeq0.0337$), consistently with the redshift of the cluster. As visible in Fig.\ref{fig:HSTzoom}, these sources appear quite blue (the SFR of the entire region is $ \sim500 M_{\odot}/yr$, \citealt{Santos15}) with disturbed morphology, and seems to be clustered in two galaxy complexes:
\begin{itemize}
\item \textbf{Complex A}, in the top-left corner of the central region of XDCP0044, includes at least 4 galaxies within 20 kpc in projected distance. It was detected in the HAWK-I image by \citet{Fassbender14} as a single source and identified as the BCG, although with several extensions, interpreted as sign of ongoing or recent mergers. 
\item \textbf{Complex B}, located at the center of the analyzed field, is made of two sources at $ d_{proj}\sim5 kpc$. One of the sources is an X-ray AGN discovered by \citet{Tozzi13}
\end{itemize}

The two complexes are very close to each other, i.e. $d_{proj}\sim35 kpc$. 

\subsection{AGN and SF activity}

Interestingly, in two of the analyzed sources, \texttt{a2} at the center of Complex A and \texttt{b1} in Complex B, the H$\alpha$ emission line is broad (FWHM$>$1500 km/s), indicating galaxies with active nuclei (AGN).  Moreover, the analysis of the line ratios ([NII]/H$\alpha$ vs [OIII]/H$\beta$) in the BPT diagram for \texttt{a3} (the only source with all four lines measured), showed that  \texttt{a3} is indeed an AGN. Three AGN have been thus discovered in the central (very small, i.e. $\sim70kpc\times70kpc$) region of the cluster (red and blue stars in the right panel of Fig. \ref{fig:HSTzoom}).
Source \texttt{b1}, at the center of Complex B,  is an X-ray point like source. From the analysis of its Chandra X-ray spectrum, we found that it is a luminous ($ L_{[2-10keV]}\sim10^{44} erg/s$) and moderately obscured ($log(N_H/cm^2)=22.7$) AGN. 
From the broad (FWHM$\sim$2200 km/s) H$\alpha$ line, we estimated its BH mass using the virial formula by \citet{GreeneHo}, finding $M_{BH} = 7.2 \times 10^7 M_{\odot}$. Moreover, the bolometric luminosity has been computed by applying the bolometric correction by \citet{Runnoe12} to the 5100\AA\ luminosity, estimated from the linear interpolation of the F105W and F140W HST magnitudes. 
Combining the derived parameters, we found that \texttt{b1} is accreting at a high rate, i.e. $\lambda_{Edd}=0.46$. 
Source \texttt{b1} has also 
been detected in ALMA continuum at 230 GHz which revealed cold dust emission from the host galaxy, detected at 5$\sigma$ significance. Assuming different QSO SEDs and normalizying them at the observed ALMA flux, we derived a SFR in the range $[150 - 490] M_{\odot}/yr$, consistent with the SFR derived from Herschel for the entire central region \citep[$452\pm58 M_{\odot}/yr$,][]{Santos15}, suggesting that most of the IR emission, and therefore of the SF, might be associated to \texttt{b1}. 
No radio emission is associated to this source. 
Finally, we derived the stellar mass of its host galaxy assuming a constant $M/L_K$ ratio \citep{Madau98} and a \citet{Chabrier03} initial mass function, finding $logM_{\star}\sim11.8 M_{\odot}$.  

Sources \texttt{a2} and \texttt{a3}, in the very crowded Complex A, have not been detected in the Chandra X-ray data. A 3$\sigma$ upper limit on the X-ray luminosity has been derived to be of the order of $L_{[2-10keV]}<10^{43} erg/s$, assuming an unabsorbed power-law with $\Gamma$=1.9. For \texttt{a2}, from the bolometric luminosity, as derived from L(5100\AA)  (i.e. $ log(L_{bol}/[erg/s]) \sim45.4$), and by applying $k_{bol}[2-10keV]\sim18.96$ \citep{Duras20}, we expect an intrinsic luminosity  $L_{[2-10keV]}\sim10^{44} erg/s$. Such value is more than 1dex higher compared to the derived X-ray luminosity upper limit, thus suggesting a high level of X-ray obscuration ($log(N_H/cm^2)>23.8$). From the broad ($FWHM\sim1900 km/s$) $H\alpha$ line, we estimated that the  BH of \texttt{a2} has a mass of $M_{BH}=3.2\times10^7 M_{\odot}$, and accretes at $\lambda_{Edd}\sim0.51$. 
While no emission has been found in ALMA corresponding to \texttt{a2},  JVLA 1.5 GHz extended radio emission has been detected spatially correlated to it. Under the assumption that such radio signal is produced by a single source, its power (logP[1.5 GHz] = 23.45) would suggest a likely (60 to 80\% of probability)
AGN powered radio emission, according to the relation introduced by \citet{Magliocchetti14,Magliocchetti18}. However, there is a not negligible probability that such  emission is on the contrary due to SF processes. In this case the measured radio luminosity would translate into a $SFR\sim100~M_{\odot}/yr$ according to the relation by \citet{Brown17}.
Finally, as for \texttt{b1}, we derived the stellar mass of its host galaxy, finding $logM_{\star}\sim11.5 M_{\odot}$. \\
Summarizing, in the core of XDCP0044 we discovered three AGN hosted in massive and star-forming galaxies, i.e. two luminous, highly accreting and obscured/moderately obscured type-1 AGN and one X-ray and optically obscured type-2 AGN.

\subsection{The $M_{BH}-M_{\star}$ plane}
\begin{SCfigure}[0.8]
\centering
\caption{\footnotesize{Correlations between central  black hole mass and host galaxy total stellar mass for the two AGN discovered in the core of XDCP004 (black circles with errors). As reference, the dashed red line is a linear fit to the sample of \citet[][KH13]{Kormendy13} by \citet{Shankar19b}, while the solid green line is the fit to  local AGN by \citet[][RV15]{Reines15} and the short-dashed blu line is  the fit by  \citet{Suh20} to local + high-z AGN. Finally the solid black line  is the de-biased $ M_{BH}-M_{\star}$ relation derived by \citet[][S+16]{Shankar16} with its scatter (yellow area).}
\vspace{2.7cm}}
\includegraphics[width=8cm]{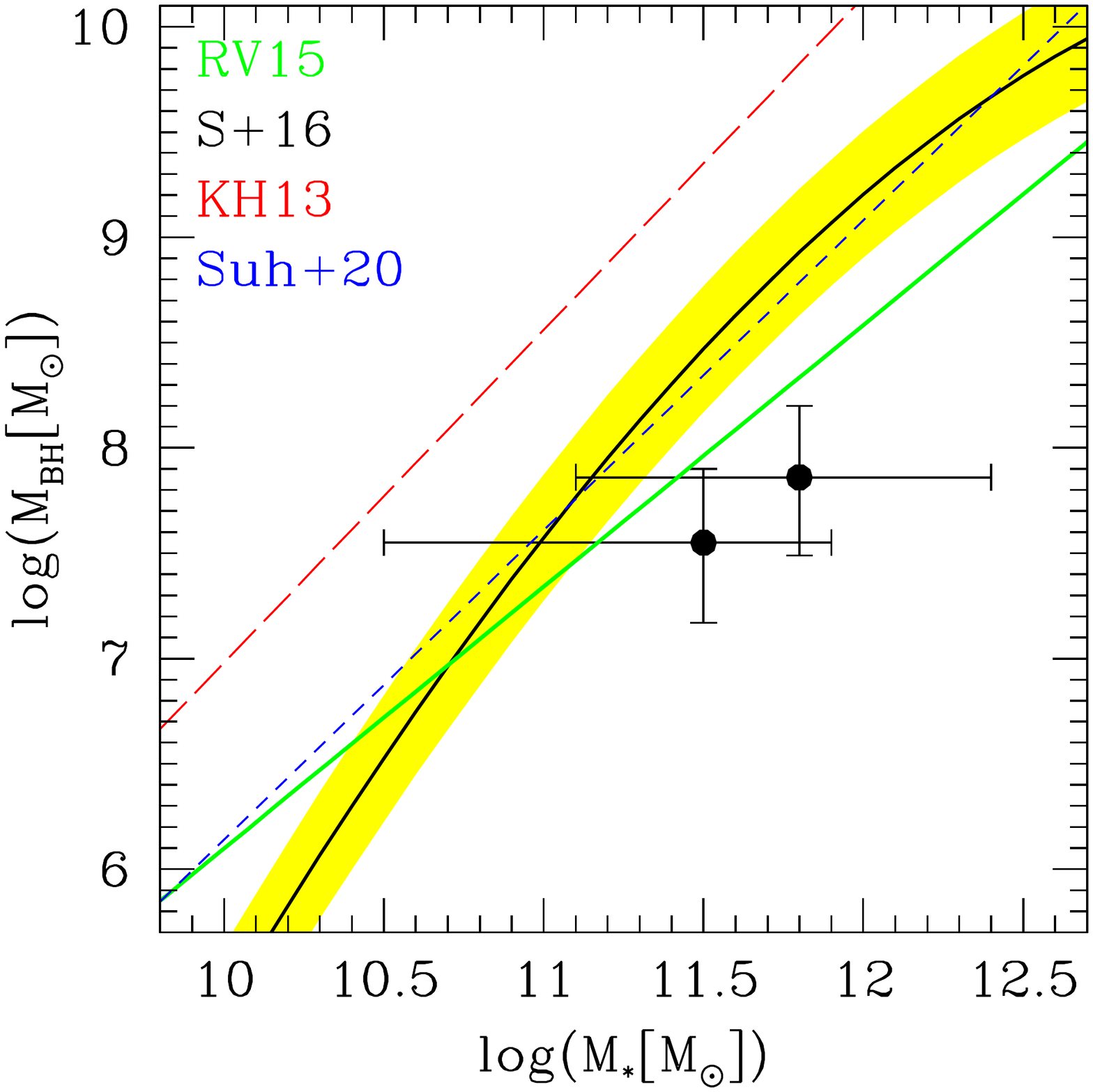}
\label{fig:scaling}
\end{SCfigure}

We studied the location of the two discovered type-1 AGN with broad H$\alpha$ in the $M_{BH}-M_{\star}$ plane (Fig. \ref{fig:scaling}).  Both of them lie below the \citet{Kormendy13} relation for local inactive galaxies with $\Delta log(M_{BH}/M_{\star})$,  computed perpendicular to the local relation, at $\sim$ 2$\sigma$ from it.
Their location is more consistent, within the large errors, with the more recent determination of the local scaling relation for active galaxies at $z<0.055$ computed by \citet{Reines15} (green line in Fig. \ref{fig:scaling}). Moreover, our data points lie close to the fit  recently found by \citet{Suh20} by including local plus high-z (up to z=2.5) AGN and to the unbiased $M_{BH}-M_{\star}$ relation computed by \citet{Shankar16}, who interpreted the discrepancy between the observed location of quiescient and active galaxies in the $M_{BH}-M_{\star}$  plane as an observational bias \citep{Shankar19b}. Indeed, our newly discovered $z\sim1.6$ AGN have $M_{BH}/M_{\star}$ ratio consistent with local active galaxies, thus showing no or negligible evolution in the intrinsic $M_{BH}-M_{\star}$ relation, in agreement with most recent works \citep{Shankar19b, Suh20}.

\section{Conclusions}

XDCP0044, a massive galaxy cluster at $z\sim1.6$, allowed us to study the processes responsible for the BCG formation in the epoch when both SF and
nuclear activity are at their peak \citep{Madau14}. 
We confirm that high-z galaxy cluster cores show different properties compared to the z=0 ones. Indeed, no single, early-type BCG has been detected in the core of XDCP0044, which is found to host a large number (at least 7 confirmed) of highly star-forming interacting galaxies, grouped in two main merging systems, both hosting also AGN activity. These three AGN provide one of the closest AGN triple at z$>$1 revealed so far with a minimum (maximum) projected distance of 10 kpc (40 kpc) and their proximity implies a future merger between them. 

These results lead to a scenario in which the AGN activity is triggered during the formation of the cluster BCG, when mergers between gas-rich galaxies provide the fuel for the AGN and for triggering starburst activity in galactic nuclei. 
Assuming that the 7 confirmed cluster members will merge to form the local BCG, we find that in a time scale of a couple of Gyrs, all galaxies in the core of XDCP0044 will experience several major mergers, forming a massive central galaxy with a final stellar mass of $1.0 \times10^{12}M_{\odot}$ at z$\sim$1, in agreement with what predicted by semi-analytic models \citep{DeLucia2007}.
In fact, we considered the case in which Complex A and B are cluster’s sub-clumps, each of which will aggregate to form a cD-like galaxy through a gravitational phase transition and then move towards the X-ray centroid in a dynamical friction time to finally merge to form the final BCG. According to this scenario, all these galaxies will merge in $\sim$2.5 Gyrs. 



\end{document}